# Free-Space Propagation and Skyrmion Topology of Toroidal Electromagnetic Pulses

Ren Wang[1,2*], Zhi-Qiang Hu[1], Pan-Yi Bao[1], Shuai Shi[1], Bing-Zhong Wang[1], Nikolay I. Zheludev[3,4,5]
and Yijie Shen[3,4*]

[1] *Institute of Applied Physics, University of Electronic Science and Technology of China, Chengdu 611731, China*
[2] *Yangtze Delta Region Institute (Huzhou), University of Electronic Science and Technology of China, Huzhou 313098, China*
[3] *Division of Physics and Applied Physics, School of Physical and Mathematical Sciences, Nanyang Technological University, Singapore 637378, Singapore*
[4] *Centre for Disruptive Photonic Technologies, The Photonics Institute, Nanyang Technological University, Singapore 637378, Singapore*
[5] *Optoelectronics Research Centre & Centre for Photonic Metamaterials, University of Southampton, Southampton, UK*

* e-mail: rwang@uestc.edu.cn and yijie.shen@ntu.edu.sg

Toroidal electromagnetic pulses have been recently reported as nontransverse, space-time nonseparable topological excitations of free space [Nat. Photon. 16, 523–528 (2022)]. However, their propagation dynamics and topological configurations have not been comprehensively experimentally characterized. Here, we report that microwave toroidal pulses can be launched by a broadband conical horn antenna. We experimentally map their skyrmionic textures and demonstrate how that during propagation the pulses evolves towards stronger space-time nonseparability and closer proximity to the canonical Hellwarth and Nouchi toroidal pulses.

Topologically structured complex electromagnetic waves have been proposed as potential information and energy carriers [1-5] for ultra-capacity communications [6,7], super-resolution metrology or microscopy [8,9] and nontrivial light-matter interactions [10,11]. Toroidal structures were recently observed in scalar spatiotemporal light waves [12], and vector electromagnetic fields termed toroidal light pulses or "flying doughnuts" [13]. Toroidal electromagnetic pulses, the propagating counterparts of localized toroidal dipole excitations in matter [14], have many exciting properties such as multiple singularities [15], space-time nonseparability [16,17] and skyrmion topologies [18,19]. Moreover, toroidal light pulses can be engaged in complex interactions with matter [20,21] and couple to electromagnetic anapoles [22] opening exiting opportunities for information and energy transfer, spectroscopy and remote sensing. However, the propagation dynamics of toroidal electromagnetic pulses and detailed characterization of their topological structures have not been experimentally investigated yet.

In this Letter, we present a new simple and robust approach to the generation of free-space microwave electromagnetic toroidal pulses with a purposely designed finite-aperture horn antenna. We experimentally study their propagation dynamics and demonstrate that during propagation the pulses launched from the finite-aperture source evolve towards higher space-time nonseparability and closer proximity to the canonical Hellwarth-Nouchi toroidal electromagnetic pulse. We undertake vectorial spatiotemporal mapping of the electric field of the pulse and demonstrate that their topological configurations are robust over a long distance.

The toroidal pulses that we will call canonical Hellwarth-Nouchi pulses are space-time nonseparable, non-transverse propagating electromagnetic excitation, the exact solution of Maxwell's equations in the form first found by Hellwarth and Nouchi in 1996 [23]:

$$E_r = 4if_0 \sqrt{\frac{\mu_0}{\varepsilon_0}} \frac{r(q_2 - q_1 - 2iz)}{[r^2 + (q_1 + i\tau)(q_2 - i\sigma)]^3}, \quad (1)$$

$$E_z = -4f_0 \sqrt{\frac{\mu_0}{\varepsilon_0}} \frac{r^2 - (q_1 + i\tau)(q_2 - i\sigma)}{[r^2 + (q_1 + i\tau)(q_2 - i\sigma)]^3}, \quad (2)$$

$$H_\theta = -4if_0 \frac{r(q_1 + q_2 - 2ict)}{[r^2 + (q_1 + i\tau)(q_2 - i\sigma)]^3}. \quad (3)$$

where $(r, z)$ represents spatial cylindrical coordinate, $t$ is time, $\sigma = z + ct$, $\tau = z - ct$, $f_0$ is a normalization constant, $q_1$ and $q_2$ represent the central wavelength of the wave package and the Rayleigh range, respectively. The magnetic field is azimuthal, $H_\theta$, and the electric field include both radial and longitudinal components, $E_r$ and $E_z$, forming a nontransverse wave. Such toroidal optical pulses wave already been observed by converting a short radially polarized pulse on a dispersive metasurface, while THz pulses were generated by optical rectification of femtosecond pulses on a nonlinear metasurface [13,24].

Our generation scheme for microwave toroidal pulses is schematically shown in Fig. 1. The generator is a radially polarized purposely designed broadband conical coaxial horn antenna with operating frequency range of 1.3-10 GHz, see details in Supplementary Materials. Below we will show results for generating toroidal pulses with $q_1 = 0.01$m and $q_2 = 50q_1$.

To launch toroidal electromagnetic pulse, the antenna was stimulated by an integral waveform as presented on Fig 1(b). As the antenna is a capacitive load to the feed, its output waveform is a differential of the driving signal that matches temporal profile of the desired free-space toroidal pulse with $q_2 = 0.5$ m and single cycle at $z = 0$ m.

Experiments were performed in a microwave anechoic chamber. To map the magnitude and phase distributions of the $E_r$ components of the broadband conical coaxial horn antenna, as shown in Fig. 1(c), we used linearly polarized horn probe with operating frequency range of 1-18 GHz. The $E_z$ component was retrieved using Gauss's law, see the details in Supplementary Materials.

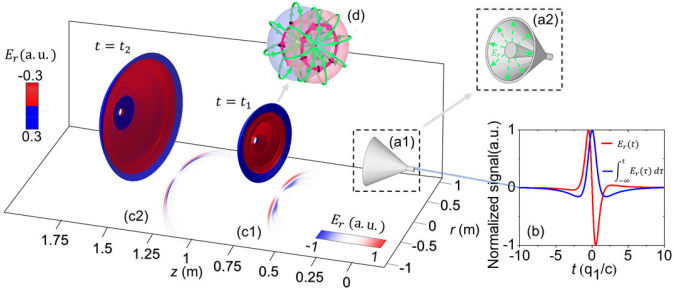

**Figure 1.** Generation of microwave toroidal pulses: (a) cylindrical coaxial antenna horn, front (a1) and back views (a2); (b) driving voltage applied to the antenna feed (blue line) and transient antenna output (red line); (c) the simulated spatiotemporal evolution of the toroidal pulse: (c1) and (c2) are the spatial isosurfaces of the electric field at two different moments of time; (d) schematic of the electromagnetic configuration of the toroidal pulse.

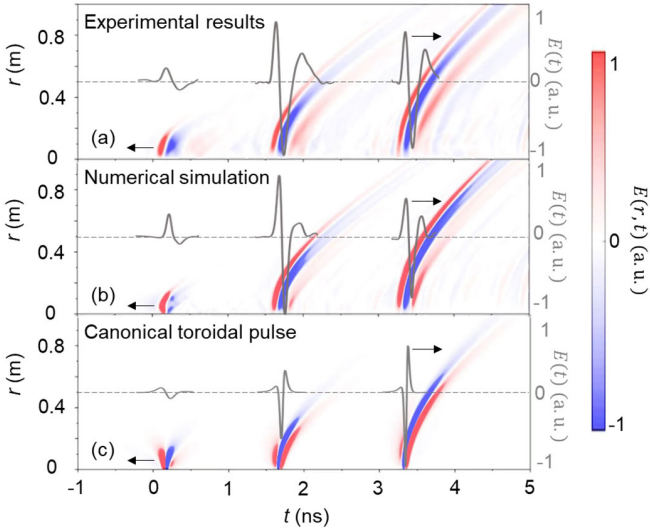

**Figure 2.** (a) Experimental and (b) numerically simulated spatiotemporal evolution of the amplitude $E_r$ of the pulses launched by the antenna compared to (c) canonical Hellwarth-Nouchi toroidal pulses. The gray curves indicate the electric field at approximately $r = 0.2$ m.

Figure 2 shows the spatiotemporal evolutions of experimentally measured, numerically simulated and canonical Hellwarth-Nouchi toroidal electromagnetic pulses at propagation distances of 5 cm, 50 cm, and 100 cm, respectively, from the horn aperture.

The conical coaxial horn antenna generates in free space an electromagnetic field of rotational symmetry around the propagation direction. From Fig. 2, it is evident that both the measured waveforms and the waveforms simulated by time-domain Maxwell solving follow a pattern similar to that of canonical Hellwarth-Nouchi pulses, transitioning from a single cycle to 1½ cycles [25]. As the bandwidth of our antenna is limited, the durations of experimentally observed pulses are somewhat larger than the canonical Hellwarth-Nouchi pulses with the same values of $q_1$ and $q_2$.

The spectral distributions at positions 5 cm, 50 cm, and 100 cm along a specific radius are depicted in Fig. 3. The spectral composition in simulated, experimental measurements, and canonical Hellwarth-Nouchi pulses spread outward with propagation and the locations of spectral maxima at different frequencies are gradually moving apart, revealing the isodiffraction characteristic [16,17].

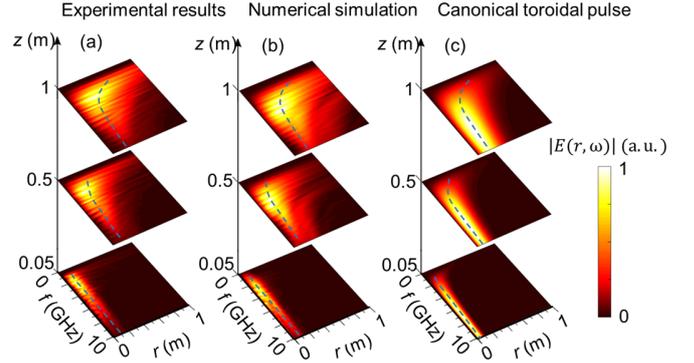

**Figure 3.** Spectral distributions of $E_r$ field at variant propagation distances: (a) experimental data, (b) numerical simulation, and (c) canonical Hellwarth-Nouchi pulses. The blue dashed lines track the spectrum maximum.

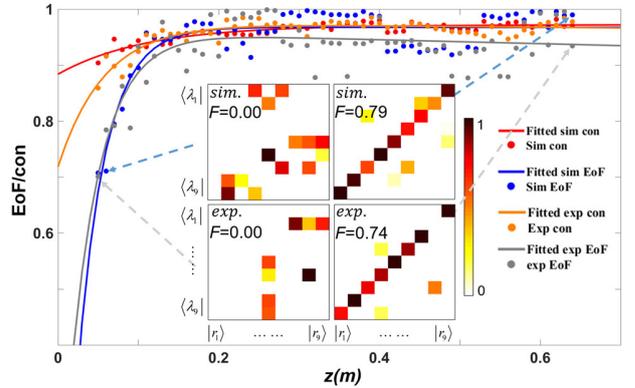

**Figure 4.** Space-time nonseparability: the numerically simulated and experimentally measured values of concurrence (con) and entanglement of formation (EoF) of the generated toroidal pulses versus propagation distance. The inserts show the state-tomography matrix for both simulation and experimental data at different distances.

The toroidal electromagnetic pulse's isodiffraction characteristic, relevant to space-time nonseparability [16], is evaluated to assess how it evolves after radiating from the antenna, Fig.4. Both simulated and measured state-tomography matrix $\{c_{i,j}\}$, with element $c_{i,j} = \int \varepsilon_{\eta_i} \varepsilon_{\lambda_j}^* dr$ representing the overlap of spatial and spectral states, indicates a poor match to the canonical Hellwarth-Nouchi pulse at proximity to the antenna and it gradually diagonalizes upon propagation, where $\varepsilon_{\lambda_j}$ and $\varepsilon_{\eta_i}$ describe the distributions of monochromatic energy density and total energy density [16]. The concurrence $con = $ $\sqrt{2[1-Tr(\rho_A^2)]}/\sqrt{2(1-1/n)}$ and entanglement of formation $EoF = -Tr[\rho_A \log_2(\rho_A)]/\log_2(n)$, where $n$ and $\rho_A$ are respectively state dimension and the reduced density matrix [24], corresponding to the simulated and measured toroidal electromagnetic pulses quickly increase and remain above 0.9 with distance. Both simulated and measured fidelity $F = Tr(M_1 M_2)$ [27], where $M_1$ and $M_2$ are respectively the density matrices for the generated and canonical Hellwarth-Nouchi toroidal pulse, at $z = 0.65$ m exceeds 0.7, indicating a high spatiotemporal nonseparability, akin to Hellwarth-Nouchi pulses with noise [16]. Therefore, during propagation the experimentally generated pulses evolves towards stronger space-time nonseparability and closer proximity to the canonical Hellwarth-Nouchi pulse.

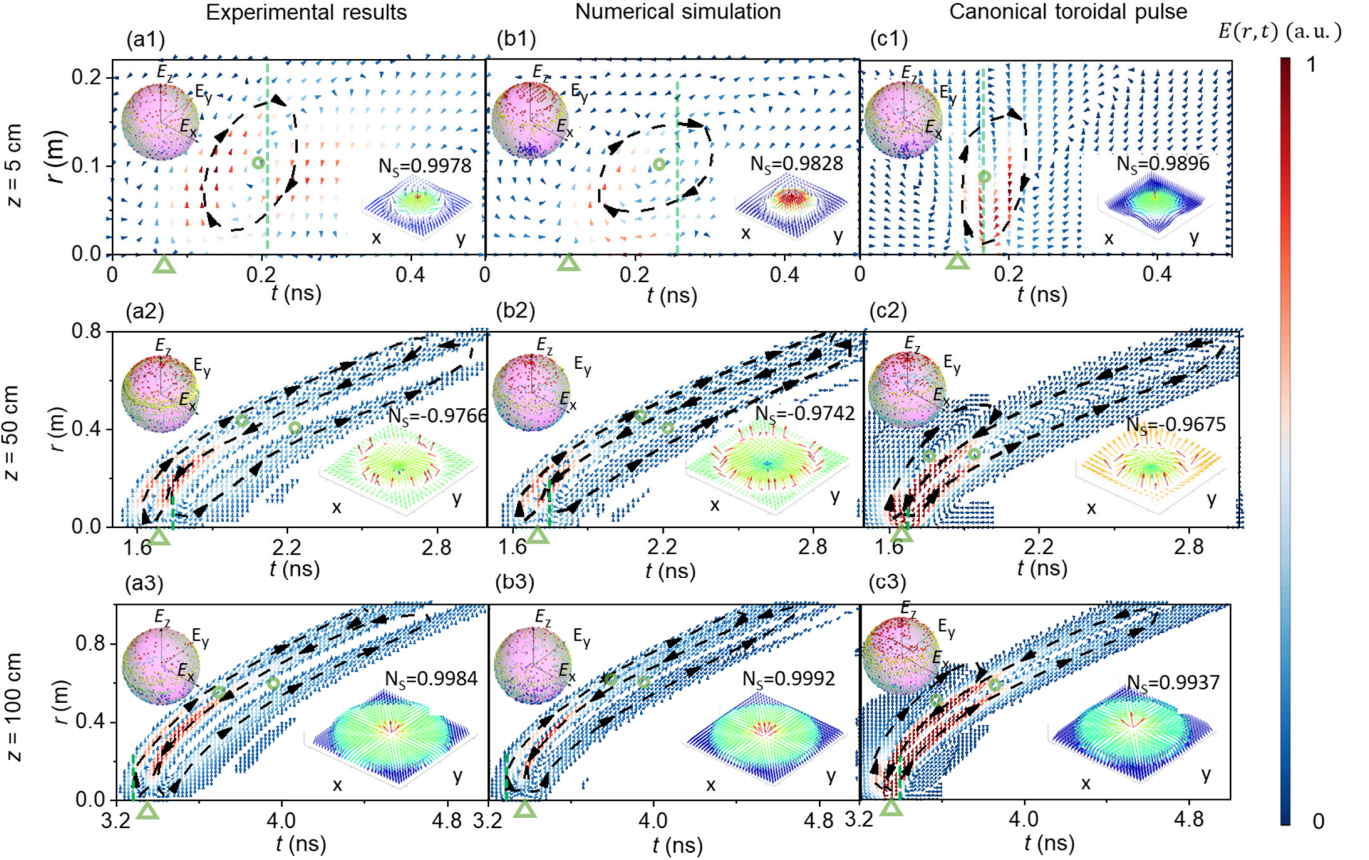

**Figure 5.** Spatiotemporal distribution of vector fields, including (a) experimental results, (b) numerical simulations, and (c) canonical Hellwarth-Nouchi toroidal pulses. (a1)~(c1), (a2)~(c2), and (a3)~(c3) correspond to the toroidal pulses propagating to $z=$ 5 cm, 50 cm, and 100 cm, respectively. In each panel, the lower-right inset features green dashed lines indicating the skyrmionic textures at a specific time on the *xy* plane. The upper-left inset corresponds to the coverage of the sphere of directions for the patterns displayed in the lower-right inset. NS represents the skyrmion number, calculated from the lower-left inset. Green tringles and circles mark the positions of saddle points and vortex rings, respectively.

Toroidal electromagnetic pulses comprise of topologically complex electromagnetic fields carrying skyrmion textures in their transverse planes, as theoretically demonstrated in Ref. [18]. However, such structures have never been observed experimentally before. Here we provide experimental mapping of these fields. Figure 5 displays maps of the vector fields, with highlighting of the measured skyrmion textures at distances of 5 cm, 50 cm, and 100 cm from the antenna aperture. As expected, the electric field has both radial component $E_r$ and longitudinal component $E_z$. The field features vector singularities, including saddle points on central axis ("longitudinal-toward radial-outward" or "radial-toward longitudinal-outward", marked by "△") and vortex rings away from the central axis (surrounding electric vector forming a vortex loop, marked by "○"). Néel-type skyrmionic textures exist in the transverse planes marked by green dashed lines, where the electric vector changes its direction from "down/up" at the center to "up/down" away from the center. Indeed, the skyrmion number of measured and simulated toroidal is always approximately ±1, as appropriate for skyrmionic textures. The coverage of the sphere of field vectors for measured, simulated, and canonical Hellwarth-Nouchi toroidal pulses fully spans the surface of the sphere, providing a confirmation of the presence of skyrmions, the calculation method of which we used is similar to that of the recent observation of sound-wave skyrmions [28], see details in Supplementary Materials.

In conclusion, we presented a simple and efficient scheme for generating microwave toroidal pulse using radially-polarized conical horn antenna. We investigated propagation of toroidal pulses and mapped their skyrmionic structure. We demonstrated that during free-space propagation the pulses evolve towards higher space-time nonseparability and closer proximity to the canonical Hellwarth-Nouchi toroidal pulses. We argue that horn antennas offer practical opportunity for using robust toroidal pulses as information carriers in high-capacity telecom applications and remote sensing.

**Acknowledgments**

This work has been supported by the National Natural Science Foundation of China (62171081), the Natural Science Foundation of Sichuan Province (2022NSFSC0039) and by European Research Council (FLEET-786851). Y. S. acknowledges the support from Nanyang Technological University via a Start Up Grant.

**Reference**

1. Nye, J. F. & Berry, M. V. Dislocations in wave trains. In A Half-Century of Physical Asymptotics and Other Di-versions: Selected Works by Michael Berry, 6–31 (World Scientific, 1974).
2. Berry, M. Making waves in physics. *Nature* **403**, 21–21 (2000).
3. He, C., Shen, Y., & Forbes, A. Towards higher-dimensional structured light. *Light: Sci. & Appl.* **11**(1) 205. (2022).
4. Bliokh, K. Y., Karimi, E., et al. Roadmap on structured waves. *J. Opt.* **25**(10), 103001 (2023).
5. Shen, Y., Zhan. Q., et al. Roadmap on spatiotemporal light fields. *J. Opt.* **25**(9), 093001 (2023).
6. Wan, Z., Wang, H., Liu, Q., Fu, X., & Shen, Y. Ultra-Degree-of-Freedom Structured Light for Ultracapacity Information Carriers. *ACS Photonics* **10**(7), 2149–2164 (2023).
7. Willner, A. E., Pang, K., Song, H., Zou, K., & Zhou, H. Orbital angular momentum of light for communications. *Appl. Phys. Rev.* **8**(4), 041312 (2021).
8. Yuan, G. H., & Zheludev, N. I. Detecting nanometric displacements with optical ruler metrology. *Science* **364**(6442), 771-775 (2019).
9. Zheludev, N. I., & Yuan, G. Optical superoscillation technologies beyond the diffraction limit. *Nat. Rev. Phys.* **4**(1), 16-32 (2022).
10. Ozawa, T. et al. Topological photonics. *Reviews of Modern Physics* **91**, 015006 (2019).
11. Price, H., Chong, Y., et al. Roadmap on topological photonics. *Journal of Physics: Photonics* **4**(3), 032501 (2022).
12. Wan, C., Cao, Q., Chen, J., Chong, A., & Zhan, Q. Toroidal vortices of light. *Nat. Photon.* **16**(7), 519-522 (2022).
13. Zdagkas, A., McDonnell, C., Deng, J., et al. Observation of toroidal pulses of light. *Nat. Photon.* **16**, 523–528 (2022).
14. Kaelberer, T., Fedotov, V. A., Papasimakis, N., Tsai, D. P., & Zheludev, N. I.. Toroidal dipolar response in a metamaterial. *Science* **330**(6010), 1510-1512 (2010).
15. Zdagkas, A., Papasimakis, N., Savinov, V., Dennis, M. R., & Zheludev, N. I.. Singularities in the flying electromagnetic doughnuts. *Nanophotonics* **8**(8), 1379-1385 (2019).
16. Shen, Y., Zdagkas, A., Papasimakis, N., & Zheludev, N. I. Measures of space-time nonseparability of electromagnetic pulses. *Phys. Rev. Res.* **3**(1), 013236 (2021).
17. Shen, Y. & Rosales-Guzmán, C. Nonseparable states of light: from quantum to classical. *Laser & Photonics Reviews* **16**(7), 2100533 (2022).
18. Shen, Y., Hou, Y., Papasimakis, N., & Zheludev, N. I. Supertoroidal light pulses as electromagnetic skyrmions propagating in free space. *Nat. Commun.* **12**(1), 5891 (2021).
19. Shen, Y., Zhang, Q., Shi, P., Du, L., Zayats, A. V., & Yuan, X. Optical skyrmions and other topological quasiparticles of light. *Nature Photonics* (2023).
20. Raybould, T., Fedotov, V. A., Papasimakis, N., Youngs, I., & Zheludev, N. I. Exciting dynamic anapoles with electromagnetic doughnut pulses. *Applied Physics Letters* **111**(8), 081104 (2017).
21. Papasimakis, N., Fedotov, V. A., Savinov, V., Raybould, T. A., & Zheludev, N. I. Electromagnetic toroidal excitations in matter and free space. *Nature Materials* **15**(3), 263-271(2016).
22. Fedotov, V. A., Rogacheva, A. V., Savinov, V., Tsai, D. P., & Zheludev, N. I.. Resonant transparency and non-trivial non-radiating excitations in toroidal metamaterials. *Sci. Rep.* **3**(1), 2967 (2013).
23. Hellwarth, R. W. & Nouchi, P. Focused one-cycle electromagnetic pulses. *Phys. Rev. E* **54**, 889–895 (1996).
24. Jana, K., Mi, Y., Møller, S. H., et al. Flying doughnut terahertz pulses generated from semiconductor currents. *arXiv preprint* arXiv:2310.06262 (2023).
25. Raybould, T., Fedotov, V., Papasimakis, N., Youngs, I., & Zheludev, N.. Focused electromagnetic doughnut pulses and their interaction with interfaces and nanostructures. *Opt. Express* **24**(4), 3150-3161 (2016).
26. P. Rungta, V. Bužek, C. M. Caves, M. Hillery, and G. J. Milburn, Universal state inversion and concurrence in arbitrary dimensions, *Phys. Rev. A* **64**, 042315 (2001).
27. D. F. V. James, P. G. Kwiat, W. J. Munro, and A. G. White, On the measurement of qubits, in Asymptotic Theory of Quantum Statistical Inference: Selected Papers (World Scientific, Singapore, 2005), pp. 509–538.
28. Muelas-Hurtado, R. D., Volke-Sepúlveda, K., Ealo, J. L., et al. Observation of polarization singularities and topological textures in sound waves. *Phys. Rev. Lett.* **129**, 204301 (2022).